\documentstyle[prl,twocolumn,aps,psfig]{revtex}
\begin{document}
\draft \twocolumn[\hsize\textwidth\columnwidth\hsize\csname
@twocolumnfalse\endcsname
\title{Strongly Enhanced Curie Temperature in Carbon-Doped Mn$_5$Ge$_3$ Films}
\author{M. Gajdzik, C. S\"urgers, M. Kelemen,  H. v. L\"ohneysen}
\address{Physikalisches Institut, Universit\"at Karlsruhe,
D-76128 Karlsruhe, Germany}
\date{\today}
\maketitle
\begin{abstract}
The structural and magnetic properties of Mn$_5$Ge$_3$C$_x$ films
prepared at elevated substrate temperatures $T_S$ are
investigated. In particular, films with $x \geq 0.5$ and $T_S$ =
680 K exhibit a strongly enhanced Curie temperature $T_{\rm C}$ =
440 K compared to Mn$_5$Ge$_3$ with $T_{\rm C}$ = 304 K.
Structural analysis of these films suggests that the carbon is
interstitially incorporated into the voids of Mn octahedra of the
hexagonal Mn$_5$Si$_3$-type structure giving rise to a lattice
compression. The enhanced ferromagnetic stability in connection
with the lattice compression is interpreted in terms of an Mn-Mn
interaction mediated by C based on a change in the electronic
structure.
\end{abstract}
\pacs{PACS numbers: 68.55.Ln, 75.50.Cc, 75.70.Ak}

]

Manganese compounds exhibit a variety of properties which are of
fundamental interest, such as the discovery of the colossal
magnetoresistance in La$_{1-x}$Sr$_x$MnO$_3$ \cite{Helmholt} or
the quantum-critical behavior of the weak itinerant magnet MnSi
\cite{Pfleiderer}. Intermetallic compounds of Mn and Ge occur in
several structural phases, which mostly exhibit antiferromagnetic
or ferrimagnetic order with rather low ordering temperatures.
However, Mn$_5$Ge$_3$ stands out as a ferromagnet with a Curie
temperature $T_{\rm C}$ = 304 K and a uniaxial magnetic anisotropy
along the $c$ axis of the hexagonal $D$8$_8$ structure (space
group P$_6/mcm$) \cite{Castelliz,Kanematsu}. The magnetic
structure of this compound as studied by neutron scattering
\cite{Ciszewski,Forsyth} reveals two Mn sublattices with different
magnetic moments resulting in an average ordered magnetic moment
of 2.6 $\mu_{\rm B}$/Mn-atom \cite{Kappel}.

In contrast, the isostructural Mn$_5$Si$_3$ compound is
antiferromagnetic with two different magnetic phases below  N\'eel
temperatures $T_{\rm N}$ = 68 K and 98 K. Consequently, a
transition from antiferromagnetic to ferromagnetic behavior was
found in Mn$_5$(Ge$_{1-y}$Si$_y$)$_3$ alloys in dependence of $y$
\cite{Kappel76,Panissod}. More remarkably, the antiferromagnetic
Mn$_5$Si$_3$ was reported to exhibit ferromagnetic order when
doped with carbon (Mn$_5$Si$_3$C$_x$), saturating at a
doping-level $x = 0.22$ with $T_{\rm C}$ = 152 K \cite{Senateur}.
The interstitial incorporation of carbon into the voids of Mn
octaedra expands the lattice slightly and the Curie temperature
$T_{\rm C}$ varies almost linearly with doping level $x$. Even
more interestingly, $T_{\rm C}$ could be enhanced up to 350 K by
incorporating carbon by simultaneous evaporation of Mn and SiC
\cite{Gajdzik96}. A shift from antiferromagnet to ferromagnet and
an ensuing $T_{\rm C}$ increase might be simply due to a lattice
expansion as suggested by the famous Bethe-Slater curve
\cite{Slater}. Indeed, the monotonic $T_{\rm C}(y)$ dependence in
Mn$_5$(Ge$_{1-y}$Si$_y$)$_3$ \cite{Kappel76} can be interpreted in
this way. On the other hand, the incorporation of carbon might
change the electronic band-structure and thus the Mn-Mn exchange.
Hence, because the role of interstitial carbon in influencing the
magnetic properties and stabilizing the ferromagnetic order in
Mn$_5$Si$_3$ is presently not clear, it is important for an
understanding to investigate the effect of C doping on the
isostructural hexagonal ferromagnetic compound Mn$_5$Ge$_3$.

In this letter we report on the magnetic and structural properties
of Mn$_5$Ge$_3$C$_x$ films with a strongly enhanced Curie
temperature $T_{\rm C} \approx 440$ K for $x \geq 0.5$ compared to
$T_{\rm C}$ = 304 K for undoped Mn$_5$Ge$_3$. This is even higher
than $T_{\rm C}$ = 330 K of Mn$_3$GeC with the cubic perovskite
structure \cite{Goodenough}. Surprisingly, we find that the
increase of $T_{\rm C}$ upon C doping is accompanied by a {\it
decrease} in the lattice constants thus demonstrating the decisive
role of carbon in altering the electronic properties.

100-nm thick Mn$_5$Ge$_3$C$_x$ films were prepared by simultaneous
$dc$- and $rf$-magnetron sputtering from elemental Mn, Ge, and C
targets (purity 99.9 \%) in Ar atmosphere onto (11\=20) sapphire
substrates at different substrate temperatures $T_S$. They were
structurally characterized by x-ray diffraction in a standard
powder diffractometer with Cu-$K_\alpha$ radiation. Magnetic
hysteresis loops were measured by the magneto-optical Kerr effect
in transverse geometry ($t$-MOKE), i.e. with the magnetic field
aligned in the film plane and perpendicular to the plane of
incidence, for temperatures $T$ = 2 - 500 K. Additionally, the
saturation magnetic moment was obtained by SQUID magnetometry at
$T$ = 5 K with the magnetic field oriented in the film plane.

\begin{figure}
\centerline{\psfig{file=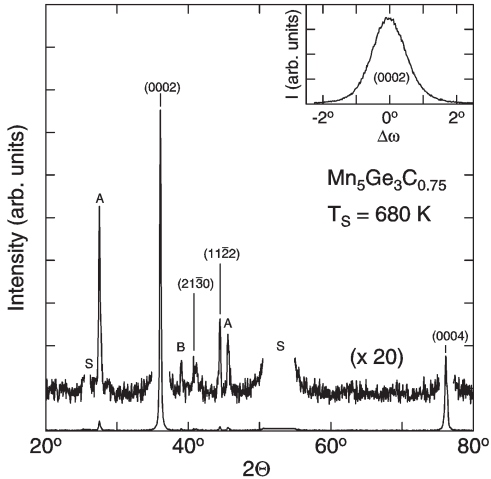,angle=-90,width=7cm,clip=}}
\caption[]{$\theta/2\theta$ x-ray diffraction pattern of an
Mn$_5$Ge$_3$C$_{0.75}$ film sputtered at $T_S$ = 680 K onto
(11\=20) sapphire (Cu $K_{\alpha}$ radiation). Miller indices
refer to the hexagonal $D8_8$ structure, see text. Reflections due
to elemental Ge, Mn$_5$Ge$_2$, and the substrate are labeled A, B,
and S, respectively. Inset shows the rocking curve ($\omega$-scan)
of the (0002) Bragg reflection.} \label{fig1}
\end{figure}

Fig. \ref{fig1} shows a $\theta/2\theta$ x-ray diffractogram of an
Mn$_5$Ge$_3$C$_{0.75}$ film prepared at $T_S$ = 680 K. The
observed diffraction lines can be indexed assuming the hexagonal
Mn$_5$Si$_3$-type structure, apart from the strong substrate
reflections and weak reflections due to precipitations of
Mn$_5$Ge$_2$ (not ferromagnetic) and elemental Ge. The structure
of such ternary $T_5M_3$C$_x$ compounds, where $T$ is a transition
metal and $M$ a metalloid, has been studied in great detail
\cite{Jeitschko,Parthe}. The atomic positions in the
Mn$_5$Si$_3$-type structure are 4 Mn$_{\rm I}$ in 4($d$), 6
Mn$_{\rm II}$ in 6($g_{\rm I}$), and 6 Si in 6($g_{\rm II}$). The
carbon is likely to be accommodated in voids surrounded by
distorted Mn$_{\rm II}$ octaedra at positions 2($b$) without a
change of the structure while a certain percentage of the Mn$_{\rm
I}$ 4($d$) positions remain unoccupied \cite{Parthe}. Such phases
with filled $D$8$_8$ structure (Nowotny phases) are generally
non-stoichiometric due to the variable occupancy of the voids.
X-ray diffractograms of samples prepared at lower and higher $T_S$
show the formation of a strongly disordered material or
precepitation of some Ge-rich alloys, respectively. The observed
intensities (Fig. \ref{fig1}) differ from the theoretically
expected values indicating a strong columnar texture along the
growth direction, i.e. the hexagonal $c$-axis, already reported
for sputtered Mn$_5$Ge$_3$ films \cite{Sawatzky}. The texture is
also inferred from the small half-width  $\Delta \omega \approx
1.2^\circ$ of the (0002) rocking curve (Fig. \ref{fig1}, inset).
We mention that the powder diffraction pattern of
Mn$_5$Si$_3$C$_x$ films prepared under the same conditions and
detached from the substrate can be perfectly fitted by the Nowotny
phase \cite{Gajdzik}. It is therefore very likely that this phase
is also formed in Mn$_5$Ge$_3$C$_x$ films.

The lattice parameters determined from the diffraction lines
indicate a lattice {\it compression} caused by the incorporation
of C. For $x = 0.75$ we find $c = 4.996$, $a = 7.135$ and $c/a$ =
0.700, i.e. a compression in each direction compared to $c =
5.053, a = 7.184, c/a = 0.703$ for Mn$_5$Ge$_3$ \cite{Castelliz}.
This is in strong contrast to Mn$_5$Si$_3$C$_x$ annealed powder
samples ($x \ll 1$), where the $c/a$ ratio increases slightly and
$a$ remains almost constant upon interstitial insertion of carbon
\cite{Parthe}.

\begin{figure}
\centerline{\psfig{file=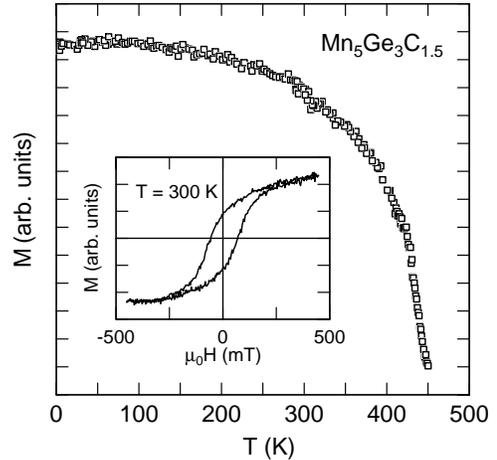,angle=-90,width=7cm,clip=}}
\caption[]{Temperature dependence of the magnetization $M$ in an
applied magnetic field $\mu _0$H = 300 mT measured by $t$-MOKE for
a Mn$_5$Ge$_3$C$_{1.5}$ film sputtered at $T_S$ = 680 K. Inset
shows a hysteresis loop $M(H)$ at room temperature.} \label{fig2}
\end{figure}

Carbon doping induces a strong enhancement of the Curie temperature,
reaching $T_{\rm C} \approx$ 440 K, i.e. well above room temperature.
Fig. \ref{fig2} shows the magnetization $M(T)$ for an Mn$_5$Ge$_3$C$_{1.5}$
film ($T_S$ = 680 K) measured by $t$-MOKE,
together with a hysteresis loop taken at room temperature.
The distinct deviation from a square loop suggests that the easy axis of
magnetization is pointing out of the film plane, i.e. along the hexagonal
$c$-axis, due to the columnar texture, since the $t$-MOKE measures the in-plane
component of the magnetization.

\begin{figure}
\centerline{\psfig{file=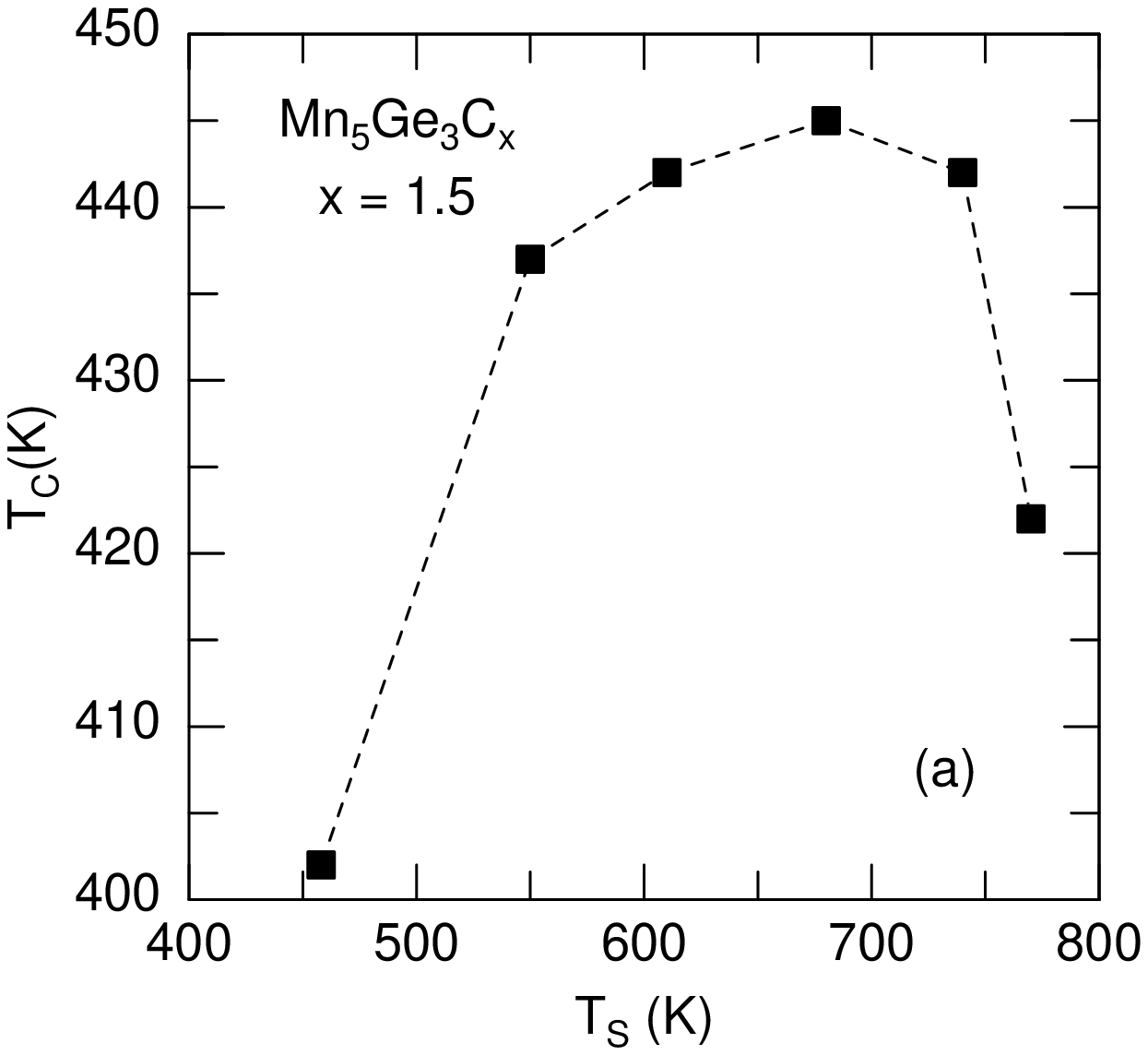,angle=-90,width=6.7cm,clip=}}
\centerline{\psfig{file=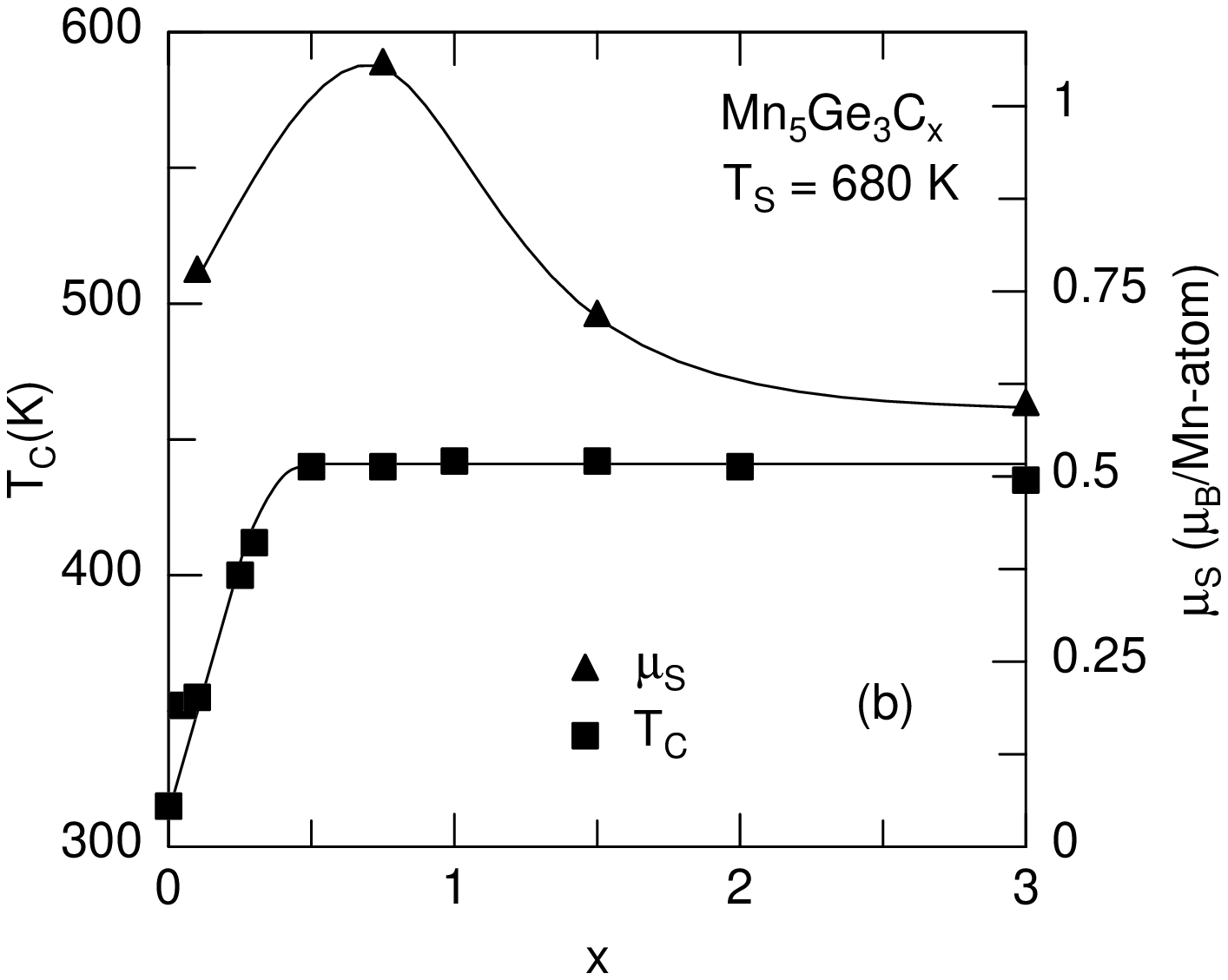,angle=-90,width=6.3cm,clip=}}
\caption[]{(a) Curie temperature $T_{\rm C}$ vs. substrate
temperature $T_S$ for sputtered Mn$_5$Ge$_3$C$_{1.5}$ films. (b)
$T_{\rm C}$ and average saturation moment $\mu_S$ in units of
$\mu_{\rm B}$/Mn-atom for Mn$_5$Ge$_3$C$_{x}$ films sputtered at
$T_S$ = 680 K vs. carbon concentration $x$. Solid lines serve as
guides to the eye.} \label{fig3}
\end{figure}

Fig. \ref{fig3}a shows $T_{\rm C}$, obtained from the extrapolation of the
$M(T)$ data to $M(T_{\rm C}) = 0$ in a small applied field $\mu_0 H$ = 5 mT,
vs. substrate temperature $T_S$.
An optimum $T_S \approx$ 680 K
can be inferred in agreement with the x-ray analysis indicating nearly
single-phase material for this $T_S$.
The formation of Mn$_5$Ge$_3$C$_x$ seems to be
thermodynamically stabilized by its low free energy together with
a high adatom mobility inherent in the growth process
at elevated temperatures.
At lower $T_S$ the process is likely to be kinetically
hindered whereas at higher $T_S$ the formation of other crystalline
phases is energetically favored.

The effect of carbon doping on the magnetic behavior was
investigated in more detail by preparing films with different $x$
at $T_S$ = 680 K. Fig. \ref{fig3}b shows $T_{\rm C}$ and the
average saturation magnetic moment $\mu_S$ plotted vs. carbon
content $x$. $T_{\rm C}$ increases linearly for small $x$ and
saturates at $x \geq 0.5$. Around $x \approx 0.5 - 0.75$, the
average magnetic moment reaches a maximum of $\mu = 1.1 \mu_{\rm
B}$/Mn-atom, considerably smaller than $\mu_S$ for Mn$_5$Ge$_3$,
and decreases for larger $x$. This indicates that the optimum C
concentration for the Mn$_5$Ge$_3$C$_x$ ferromagnetic phase is
around $x$ = 0.75. The saturation of carbon uptake is higher
compared to Mn$_5$Si$_3$C$_x$ sintered powder samples ($x_{max}$ =
0.22) \cite{Senateur} which indicates that the carbon is
distributed more homogeneously on an atomic scale during film
growth whereas in the powder samples this process is presumably
diffusion limited. This may stimulate further investigations with
the aim at incorporating even higher carbon concentrations by
employing other preparation techniques far from equilibrium.

From neutron-diffraction data, Forsyth and Brown presented
evidence that in several Mn-metalloid binary intermetallic
compounds the moment reduction at different Mn-sites with respect
to the free-ion value is governed by the nearest-neighbor Mn-Mn
interaction. Below a critical average Mn-Mn bond length of about
3.1 \AA, each Mn neighbor reduces the moment of the coordinated Mn
atom from the ionic value by about $2 \mu_{\rm B}$/\AA\,
\cite{Forsyth}. Applying this empirical relationship to the
present samples, a magnetic moment of about 2.3 $\mu_{\rm B}$ is
expected from the measured lattice parameters. The additional
reduction observed in the present samples can have two reasons.
First, it could be due to some non-magnetic minority phases.
However, this would require a high volume fraction of 50 \% of
non-magnetic material at optimal doping $x \approx 0.75$ which
seems unlikely in view of the weak parasitic x-ray Bragg
reflections that could be attributed to Mn$_5$Ge$_2$ or elemental
Ge. Second, if the carbon participates in the Mn-Mn bonding the
simple relationship between average Mn-Mn bond length and magnetic
moment will break down, because the $d$-band population and
therefore $\mu_S$ is modified by the incorporation of carbon into
the magnetic host lattice. Such an effect of carbon doping on the
band structure has been suggested for the cubic perovskites
Mn$_3M$C where $M$ is a non-transition metal \cite{Goodenough}.
Clearly, the magnetic moment at the different Mn sites has to be
determined by further microscopic investigations.

For larger concentrations $x > 0.75$ the
average Mn moment decreases whereas $T_{\rm C}$
remains constant. This behavior is attributed to the increased formation of
non-magnetic phases which reduce the amount of the ferromagnetic phase and
therefore the average moment, whereas the Curie temperature of the majority
phase is still maintained. This is corroborated by an overall
decrease of the magnetic signal in $t$-MOKE with increasing $x$.
Furthermore, the half-width of the (0002) rocking curve increases with
increasing $x$, most likely due to the decreasing columnarity of the
multiphase material.

The question arises how the ferromagnetic exchange interaction can
be enhanced by the incorporation of carbon. Obviously, only the
Mn$_{\rm II}$ atoms of the octahedra around the C site are
affected via {\it sp}-{\it d} interaction or hybridization. This
can also be inferred from neutron scattering data on
Mn$_5$(Ge$_{1-y}$Si$_y$)$_3$ \cite{Panissod}. The Mn$_{\rm II}$
atoms with a moment of $\approx 3 \mu_{\rm B}$ seem to be strongly
ferromagnetically coupled to each other while the coupling between
the Mn$_{\rm I}$ atoms with $\mu \approx 2 \mu_{\rm B}$ is
certainly weaker. Substitution of Si for Ge gives rise to a
reduction of the moment at the Mn$_{\rm I}$ site whereas the
Mn$_{\rm II}$ moments remain unaffected.

In C-doped Mn$_5$Ge$_3$ two alternative possibilities may lead to
the strongly enhanced Curie temperature, involving either a
predominantly covalent bonding between Mn and C or an ionic charge
transfer. The carbon and the surrounding Mn$_{\rm II}$ atoms are
possibly covalently bonded by {\it sp}-{\it d} hybrid orbitals and
the local environment of the nearest Mn$_{\rm II}$ neighbors
around C is comparable to the cubic Mn$_3/M$ perovskites, where C
is incorporated in octahedral interstices of the face-centered
cubic $Mn/$M-atom array \cite{Goodenough}. In the perovskites the
band structure is modified by the electronegative interstices so
that the broad bands arising from the Mn-$M$ bonding are split by
the changed translational symmetry and the occupation of the $sp$
states overlapping with the $d$ bands is changed by Mn-C and $M$-C
interactions. This may stabilize a spontaneous magnetization
although the spin configuration can be complex giving rise to a
ferromagnetic-to-antiferromagnetic transition at low temperatures
\cite{Goodenough}. For the present samples such band-structure
effects are further complicated by the variable filling of the
$(4d)$ Mn$_{\rm I}$ positions of the Nowotny phase. However, the
general mechanism of changing the band-structure due to the
introduction of C serving as an electron acceptor might be similar
to Mn$_3M$C.

On the other hand, in a simple ionic picture applied to undoped
Mn$_5$Ge$_3$, charge balance requires 6 Ge$^{4-}$, 4 Mn$_{\rm
I}^{3+}$, and 6 Mn$_{\rm II}^{2+}$. The Mn$_{\rm II}$ valence is
also compatible with the magnetization density measured by neutron
diffraction \cite{Forsyth}. The addition of C will induce a charge
transfer from Mn$_{\rm II}$ atoms to C, assuming that C is
strongly electronegative. This would, for instance, lead to a
C$^{-4}$ ion surrounded by 4 Mn$_{\rm II}^{3+}$ and 2 Mn$_{\rm
II}^{2+}$ ions at the corners of the octahedron, thus offering the
possibility of double exchange leading to ferromagnetic order as
originally proposed by Zener \cite{Zener} and investigated in
detail in the colossal magnetoresistance alloys \cite{col}. For
ferromagnetic alloys of Mn or Fe the effect of interstitial
non-magnetic atoms like B, C, and N, on the magnetic properties
has been recently studied theoretically by assuming a
Hubbard-split band at half filling coupled with a localized spin
system via exchange coupling \cite{Sakuma}. The coupling gives
rise to the double-exchange mechanism in the limit of completely
localized orbitals. Whether this model or other more realistic
double-exchange models including lattice distortions can be
applied to the C-doped Mn$_5$Ge$_3$ alloys requires a detailed
knowledge of the valences at the different Mn$_{\rm II}$ sites. In
case of a complete charge transfer we would expect a change of the
average valence from 2.4 for Mn$_5$Ge$_3$ to 3 for
Mn$_5$Ge$_3$C$_{0.75}$. We have checked this possibility by
performing preliminary x-ray absorption measurements for $x=0$ and
$x=0.75$. The experiments were done at the beamline A1 at HASYLAB,
Hamburg. The absorption edge shifts to higher energies in the
C-doped sample by less than 0.5 eV which corresponds \cite{Apte}
to an average valence change of less than 0.1, clearly much less
than the ionic model would predict. We conclude that the double
exchange mechanism plays only a minor role in enhancing the
ferromagnetism in Mn$_5$Ge$_3$C$_x$.

Although the facts that ($i$) d$T_{\rm C}$/d$p > 0$ for pure
Mn$_5$Ge$_3$ \cite{Sevast} and ($ii$) $T_{\rm C}$ increases while
the volume decreases upon C doping appear to be superficially
compatible, it is clear from the above discussion that the
enhanced Curie temperature of Mn$_5$Ge$_3$C$_x$ is not due to a
simple volume effect. This is supported by the monotonic $T_{\rm
C}(y)$ dependence in Mn$_5$(Ge$_{1-y}$Si$_y$)$_3$ as soon as the
antiferromagnetic Mn$_5$Si$_3$ changes to ferromagnetism with
increasing Ge concentration. We also note that taking d$T_{\rm
C}$/d$p$ = 5.5 K/GPa for Mn$_5$Ge$_3$ \cite{Ido} and assuming a
bulk modulus of $K$ = 110 GPa \cite{Istomina} would lead to
d$T_{\rm C}$/d$(\Delta V/V) = -605$ K which is incompatible with
d$T_{\rm C}$/d$(\Delta V/V) = -1700$ K derived by simply relating
the volume change upon C doping to the $T_{\rm C}$ change. This
strongly confirms our above assignment of a predominantly
electronic effect.

In conclusion, carbon doping of Mn$_5$Ge$_3$ compounds leads
compression of the crystalline lattice and to a strongly enhanced
ferromagnetic stability in remarkable contrast to
Mn$_5$(Ge$_{1-y}$Si$_y$)$_3$ compounds, where $T_{\rm C}$
decreases with decreasing lattice parameters. We have presented
evidence that this is due to an increased interaction between Mn
atoms mediated by interstitially incorporated carbon. The enhanced
Curie temperature is attributed to a change in the electronic band
structure. Our results have opened the route to a number of
important investigations for the future. First of all, neutron
studies are required to determine the magnetic moments at the
different Mn sites. Electron-loss spectroscopy can determine the
valence of carbon. Finally, the C bonding in these Mn compounds
should be investigated theoretically.

We thank K. Attenkofer at the HASYLAB, Hamburg, for performing the
x-ray absorption experiments and E. Dormann for useful
discussions. This work was supported by the Deutsche
Forschungsgemeinschaft.

\end{document}